\documentclass[%
 reprint,
 amsmath,amssymb,
 aps,
]{revtex4-1}
\usepackage{siunitx}
\usepackage{graphicx}
\usepackage{dcolumn}
\usepackage{bm}
\usepackage{color}
\usepackage{verbatim}
\usepackage[T1]{fontenc}
\usepackage[titletoc]{appendix}
\usepackage{subfigure}
\usepackage{float}

\begin{document}

\preprint{APS/123-QED}

\title{Analytic formulas for the D-mode Robinson instability}

\author{Tianlong He}\email{htlong@ustc.edu.cn}
\author{Weiwei Li}
\author{Zhenghe Bai}
\author{Weimin Li}

\affiliation{National Synchrotron Radiation Laboratory, University of Science and Technology of China, Hefei, Anhui, 230029, China}%

\date{\today}

\begin{abstract}
The passive superconducting harmonic cavity (PSHC) scheme is adopted by several existing and future synchrotron light source storage rings, as it has a relatively smaller $R/Q$ and a relatively larger quality factor ($Q$), which can effectively reduce the beam-loading effect and suppress the mode-one instability. Based on the mode-zero Robinson instability equation of uniformly filled rigid bunches and a search algorithm for minimum, we have revealed that the PSHC fundamental mode with a large loaded-$Q$ possibly triggers the D-mode Robinson instability [T. He, et al., Mode-zero Robinson instability in the presence of passive superconducting harmonic cavities, PRAB 26, 064403 (2023)]. This D-mode Robinson instability is unique because it is anti-damped by the radiation-damping effect. In this paper, analytical formulas for the frequency and growth rate of the D-mode Robinson instability are derived with several appropriate approximations. These analytical formulas will facilitate analyzing and understanding the D-mode Robinson instability. Most importantly, useful formulas for the D-mode threshold detuning calculation have finally been found.
\begin{description}
	
\item[PACS numbers]
29.27.Bd, 41.75.Ht
\end{description}
\end{abstract}

\pacs{Valid PACS appear here}
\maketitle


\section{Introduction}
In the 3rd generation synchrotron light sources, such as SLS and ELETTRA, passive superconducting harmonic cavities (PSHCs) have been successfully used to elongate the bunch length and thus improve the beam lifetime, with a history of over 20 years. It was reported that SLS and ELETTRA successfully increased the beam lifetime by more than a factor of 2 and 3, respectively~\cite{Pedrozzi01}. Recently, SSRF has also installed a two-cell PSHC and tested the bunch lengthening performance, successfully doubling the beam lifetime~\cite{Zhang02}. Their success has to some extent driven more 4th generation synchrotron light sources to adopt the PSHC scheme, such as HALF~\cite{Wei03}, Diamond-II~\cite{Diamond04}, and Sirius~\cite{Sirius05}, etc., expecting to lengthen the beam by a factor of at least three to ensure their high performance of operation.

The loaded quality factor ($Q$) of PSHC is generally at the level of $1\times10^8$, which is significantly different from that of the normal-conducting cavity. In a previous study~\cite{He06}, we revealed that the PSHC fundamental mode may cause a special mode-zero Robinson instability, called the D-mode Robinson instability. This instability is enhanced with a larger loaded $Q$ value and a smaller radiation damping time. The latter characteristic of the D mode is opposite to any conventional instability that can be mitigated through radiation damping. Generally, the D-mode instability will not be triggered unless the PSHC detuning is sufficiently low. That means it will impose a limitation to the bunch lengthening at relatively low currents.

This D-mode oscillation was not only observed in simulations~\cite{He06,Carmignani07,Gamelin08,Stingelin09} but is also likely to have been observed in experiments. As early as twenty years ago, in Ref.~\cite{Pedrozzi01}, it was reported, for ELETTRA that the voltage feedback loop acting on the PSHC tuning system is opened at about $\SI{160}{mA}$ because, for lower currents, an instability is then observed causing beam loss. Because $V_h\propto I_0 / \Delta f_r$, (with $V_h$, $I_0$, $\Delta f_r$ denoting the PSHC voltage, beam current, and PSHC detuning, respectively), the voltage feedback will reduce the detuning in an equal proportion as the beam current decreases. It will be shown later that the D-mode threshold current of ELETTRA is exactly calculated to be about $\SI{160}{mA}$ if the PSHC voltage is kept to meet the so called near-optimum bunch lengthening condition. Therefore, in the view of this paper, below $\SI{160}{mA}$, the PSHC detuning of ELETTRA possibly touches the D-mode detuning threshold, resulting in beam loss. Another experiment that can be treated as a possible evidence of D mode was conducted in SLS and reported at the workshop of HarmonLIP 2022. The threshold detuning for beam loss was measured to be $\SI{23.6}{kHz}$ at $\SI{100}{mA}$~\cite{Stingelin09}, which is very close to that of the theoretical D-mode threshold detuning~\cite{He06}. We also noticed that SSRF achieved a bunch lengthening by a factor of about 2.1 for the case of four uniformly distributed subtrains filling at $\SI{200}{mA}$~\cite{Zhang02}, which was still considerably lower than the theoretical near-optimum lengthening ratio. Beam loss was also observed when trying to reduce the PSHC detuning for a larger bunch lengthening ratio~\cite{Hou10}. The above resulting experiments for SSRF are consistent with the prediction based on the D-mode theory, proposed in our previous study~\cite{He06}. 

To further prove our viewpoint, we continue this study for the D-mode instability, as a follow-up to our previous study. It is found that an analytical formula for the D mode can be derived by taking appropriate approximations, and an approximate formula for calculating the D-mode threshold detuning can be further obtained. We will see the limitations imposed by the D mode to the bunch lengthening for the aforementioned three light sources at different beam currents.
 
The rest of this paper is organized as follows: In Sec.~\ref{sec:level2}, We briefly review the mode-zero Robinson instability equation, which is extended to include the radiation damping term. In Sec.~\ref{sec:level3}, the analytical formulas for the D mode are derived in detail and verified accurately using the Hefei Advanced Light Facility storage ring parameters. In Sec.~\ref{sec:level4}, we further introduce a method that can be used to obtain the D-mode threshold detuning and apply it to the evaluation of bunch lengthening limitation for three existing storage rings employed PSHC. Finally, we conclude this paper in Sec.~\ref{sec:level5}.

\section{\label{sec:level2} Mode-zero Robinson instability}
In electron storage rings uniformly filled with $M$ equal bunches, the beam instability caused by a narrowband resonator impedance can be described by a well-known equation~\cite{Chao11,KyNg12}:
\begin{equation}\label{eq:labe1}
   \begin{aligned}
   \Omega^2-\omega_{s0}^2 = -i\frac{\omega_0 I_0 \alpha_c}{2\pi E / e}\sum\limits^{\infty}_{p=-\infty}\{pM \omega_0 Z(pM \omega_0)-\\
   (pM \omega_0+\mu \omega_0+\Omega)Z(pM \omega_0+\mu \omega_0+\Omega)\},
    \end{aligned}
\end{equation}
where $\Omega=\Omega_r + i\Omega_i$ is the complex angular oscillation frequency, $\Omega_r$ represents the coherent angular frequency, $\Omega_i$ is the instability growth rate ($\Omega_i<0$ means damping), $i$ is the imaginary unit, $\omega_{s0}$ is the unperturbed synchronous angular frequency given by $\omega_{s0}=\sqrt{eV_{rf} h\omega_0 \alpha_c |\cos\varphi_s|/(ET_0)}$, $V_{rf}$ is the main voltage, $h$ is the harmonic number, $\omega_0$ is the angular revolution frequency, $\alpha_c$ is the momentum compaction factor, $\varphi_s$ is the unperturbed synchronous phase determined by $V_{rf}\sin \varphi_s=U_0$, $U_0$ is the energy loss per turn, $T_0$ is the revolution time, $E$ is the beam energy, $I_0$ is the beam average current, $\mu$ is the coupled-bunch mode number taken from $0$ to $M-1$, and $p$ is an integer taken from $-\infty$ to $+\infty$.

It should be noted that Eq.~(\ref{eq:labe1}) is derived based on the point-like-bunch and linear-wake-force model. Generally, it works well for the case of a narrowband resonator and dipole-oscillation approximation. Here, we focus only on the mode-zero ($\mu=0$) coupled bunch instability driven by the PSHC fundamental mode, whose impedance form is
\begin{equation}\label{eq:labe2}
   Z(\omega)=\frac{R}{1+iQ(\frac{\omega_r}{\omega}-\frac{\omega}{\omega_r})},
\end{equation}
where $R$, $Q$, $\omega_r$ are the characteristic parameters of a resonator, representing the shunt impedance, quality factor, and angular resonant frequency, respectively. For the mode-zero instability, the summation of the series on the right-hand side of Eq.~(\ref{eq:labe1}) is dominated by the two terms of $pM=\pm nh$. Then Eq.~(\ref{eq:labe1}) can be simplified as 
\begin{equation}\label{eq:labe3}
   \begin{aligned}
   \Omega^2-\omega^2_s + i\frac{\omega_0 I_0 \alpha_c}{2\pi E/e}\{i2nh\omega_0 \mathrm{Im}[Z(nh\omega_0)]\\
   -\omega^{+}_p Z(\omega^{+}_p)-\omega^{-}_p Z(\omega^{-}_p)\}=0,
   \end{aligned}
\end{equation}
where $\omega_p^{\pm}=\pm n h \omega_0 + \Omega$. Using the approximation of $\mathrm{Im}[Z(nh\omega_0)]\approx - R\omega_r/(2Q\Delta \omega_r)$, with $\Delta \omega_r$ being the angular detuning, Eq.~(\ref{eq:labe3}) can be further simplified as
\begin{equation}\label{eq:labe4}
   \Omega^2-\omega^2_s-ic\{Z(\omega^{+}_p)-Z(\omega^{-}_p)\}=0,
\end{equation}
where we define that
\begin{equation}\label{eq:labe5}
   \omega^2_s=\omega^2_{s0}-\frac{enh I_0 \alpha_c R \omega^2_0 \omega_r}{2 \pi E Q \Delta \omega_r},
\end{equation}
and 
\begin{equation}\label{eq:labe6}
   c=\frac{enh I_0 \alpha_c \omega^2_0}{2 \pi E}.
\end{equation}
Due to the significant role of radiation damping effect in the D-mode instability [6], it is now time to add it to Eq.~(\ref{eq:labe4}). Then it gives
\begin{equation}\label{eq:labe7}
   \Omega^2+i\frac{2\Omega}{\tau_z}-\omega^2_s-ic\{Z(\omega^{+}_p)-Z(\omega^{-}_p)\}=0.
\end{equation}
The second term on the left-hand side of Eq.~(\ref{eq:labe7}) represents the radiation damping. If we define a function $g(\Omega)$, whose form is given as follows
\begin{equation}\label{eq:labe8}
   g(\Omega)=\Omega^2+i\frac{2\Omega}{\tau_z}-\omega^2_s-ic\{Z(\omega^{+}_p)-Z(\omega^{-}_p)\},
\end{equation}
the solution of Eq.~(\ref{eq:labe8}) should meet $g(\Omega)=0$. In Ref.~\cite{He06}, a search algorithm for the minimum of $|g(\Omega)|$ was proposed to obtain the solution of Eq.~(\ref{eq:labe7}). It has been shown that there are two local minima, one is labeled as S mode and another is labeled as D mode. The D-mode Robinson instability should be paid attention to in the case of PSHC with very large loaded $Q$ values. Additionally, by fixing the real part $\Omega_r$ and drawing to analyze the dependence of $\mathrm {Re}\{Z(\omega^{+}_p)-Z(\omega^{-}_p)\}$ on the imaginary part $\Omega_i$, it was found that the dependence can be fitted by a straight line near $\Omega_i=0$, that is, it can be expressed as 
\begin{equation}\label{eq:labe9}
   \mathrm {Re}\{Z(\omega^{+}_p)-Z(\omega^{-}_p)\} \approx k\Omega_i+b,
\end{equation}
where $k$ and $b$ are real coefficients as a function of $\Omega_r$.

Inspired by Eq.~(\ref{eq:labe9}), we can derive more simplified analytical formulas for the D-mode instability by taking several reasonable approximations. Now we will show the derivation details.

\section{\label{sec:level3} Derivation of analytical formulas}
\subsection{\label{sec:level3.1} Take the real part}
Substitute $\Omega=\Omega_r + i\Omega_i$ into Eq.~(\ref{eq:labe8}) and take the real part $\mathrm {Re}[g(\Omega)]$, we have 
\begin{equation}\label{eq:labe10}
   \Omega^2_r-\Omega^2_i-i\frac{2\Omega_i}{\tau_z}-\omega^2_s+c \cdot \mathrm {Im}\{Z(\omega^{+}_p)-Z(\omega^{-}_p)\}=0.
\end{equation}
For the D-mode oscillation, its $\Omega_r$ is slightly lower than the angular detuning of PSHC. In general, we can have $\Omega_r \gg \Omega_i$. Thus, the second and the third terms on the left-hand side of Eq.~(\ref{eq:labe10}) can be ignored, and it gives
\begin{equation}\label{eq:labe11}
   \Omega^2_r-\omega^2_s+c \cdot \mathrm {Im}\{Z(\omega^{+}_p)-Z(\omega^{-}_p)\}\approx 0.
\end{equation}
Let $\Delta \omega_1=\Delta \omega_r-\Omega_r>0$ and $\Delta \omega_2=\Delta \omega_r+\Omega_r \approx 2\Delta \omega_r$, then the third term on the left-hand side of Eq.~(\ref{eq:labe11}) can be transformed as follows:
\begin{equation}\label{eq:labe12}
   \begin{aligned}
   c \cdot \mathrm {Im}\{Z(\omega^{+}_p)&-Z(\omega^{-}_p)\}=c \cdot \mathrm {Im}\{Z(\omega^{+}_p)+Z(-\omega^{-}_p)\}\\
   &=c \cdot \mathrm {Im}\{\frac{R}{1+i\frac{2Q\Delta\omega_1}{\omega_r}}+\frac{R}{1+i\frac{2Q\Delta\omega_2}{\omega_r}}\}\\
   &\approx -\frac{cR\omega_r}{2Q\Delta\omega_1}-\frac{cR\omega_r}{4Q\Delta\omega_r}.
   \end{aligned}
\end{equation}
Substitute Eq.~(\ref{eq:labe12}) into Eq.~(\ref{eq:labe11}) and take $\Omega_r=\Delta\omega_r-\Delta\omega_1$, we get
\begin{equation}\label{eq:labe13}
   \Delta\omega^3_1-2\Delta\omega_r \Delta\omega^2_1+(\Delta\omega^2_r-\omega^2_s-\frac{cR\omega_r}{4Q\Delta\omega_r})\Delta\omega_1-\frac{c\omega_r R}{2Q}=0.
\end{equation}
Note that Eq.~(\ref{eq:labe13}) is a cubic equation about $\Delta\omega_1$, which is not difficult to solve. Equation~(\ref{eq:labe13}) has three solutions, among which only the one closest to zero is required.

Actually, for the D-mode oscillation, it is in general that $\Delta\omega_1 \ll \Delta\omega_r$. Hence, the first term on the left-hand side of Eq.~(\ref{eq:labe13}) can be neglected. Then we get
\begin{equation}\label{eq:labe14}
   \Delta\omega^2_1-(\frac{\Delta\omega_r}{2}-\frac{\omega^2_s}{2\Delta\omega_r}-\frac{cR\omega_r}{8Q\Delta\omega^2_r})\Delta\omega_1+\frac{c\omega_r R}{4\Delta\omega_rQ}=0.
\end{equation}
Equation~(\ref{eq:labe14}) is a quadratic equation about $\Delta\omega_1$, which can be solved analytically. If we introduce
\begin{equation}\label{eq:labe15}
   B=\frac{\Delta\omega_r}{4}-\frac{\omega^2_s}{4\Delta\omega_r}-\frac{cR\omega_r}{16Q\Delta\omega^2_r}.
\end{equation}
and
\begin{equation}\label{eq:labe16}
   C= \frac{c\omega_r R}{4\Delta\omega_r Q},
\end{equation}
then  Eq.~(\ref{eq:labe14}) has two solutions written as $B\pm \sqrt{B^2-C}$. Only the solution closest to zero is required, which is 
\begin{equation}\label{eq:labe17}
   \Delta\omega_1= B-\sqrt{B^2-C}.
\end{equation}
The D-mode angular oscillation frequency is 
\begin{equation}\label{eq:labe18}
   \Omega_r= \Delta\omega_r-\Delta\omega_1.
\end{equation}

Please note that in the above derivations, we utilized the characteristics of the D-mode instability: firstly, the loaded $Q$-value of PSHC is very large in the level of $10^{8}$, and secondly, $\Delta\omega_1 \ll \Delta\omega_r$.

\subsection{\label{sec:level3.2} Take the imaginary part}
Take the imaginary part $\mathrm {Im}[g(\Omega)]$, it gives
\begin{equation}\label{eq:labe19}
   2\Omega_r\Omega_i+\frac{2\Omega_r}{\tau_z}-c \cdot \mathrm {Re}\{Z(\omega^{+}_p)-Z(\omega^{-}_p)\}=0.
\end{equation}

Introducing $\Delta\Omega_1=\Delta\omega_r-\Omega_r-i\Omega_i=\Delta\omega_1-i\Omega_i$ and $\Delta\Omega_2=\Delta\omega_r+\Omega_r+i\Omega_i=\Delta\omega_2+i\Omega_i$, and utilizing $\mathrm {Re}\{Z(\omega^{-}_p)\}=\mathrm {Re}\{Z(-\omega^{-}_p)\}$, the third term on the left-hand side of Eq.~(\ref{eq:labe19}) can be transformed as follows:
\begin{equation}\label{eq:labe20}
   \begin{aligned}
   c \cdot \mathrm {Re}\{Z(\omega^{+}_p)&-Z(\omega^{-}_p)\} =c \cdot \mathrm {Re}\{Z(\omega^{+}_p)-Z(-\omega^{-}_p)\}\\
   &=c \cdot \mathrm {Re}\{\frac{R}{1+i\frac{2Q\Delta\Omega_1}{\omega_r}}-\frac{R}{1+i\frac{2Q\Delta\Omega_2}{\omega_r}}\}\\
   &\approx k\Omega_i+b.
   \end{aligned}
\end{equation}
To obtain $b$, let $\Omega_i=0$ and $c \cdot \mathrm {Re}\{Z(\omega^{-}_p)\}$ can be neglected due to $\Delta\omega_2 \gg \Delta\omega_1$, then we get
\begin{equation}\label{eq:labe21}
   b|_{\Omega_i=0}\approx c \cdot \mathrm {Re} \{\frac{R}{1+i\frac{2Q\Delta\omega_1}{\omega_r}}\} \approx \frac{cR\omega^2_r}{4Q^2\Delta\omega^2_1}.
\end{equation}
To obtain $k$, we should first get the first-order derivation of $\mathrm {Re}\{Z(\omega^{+}_p)-Z(\omega^{-}_p)\}$ about $\Omega_i$, then let $\Omega_i=0$ and neglect the term related to $\mathrm {Re}\{Z(\omega^{-}_p)\}$, finally, we get
\begin{equation}\label{eq:labe22}
   k|_{\Omega_i=0}\approx -\frac{2cQ}{\omega_r R} \mathrm {Re} \{\frac{R^2}{(1+i\frac{2Q\Delta\omega_1}{\omega_r})^2}\} \approx \frac{cR\omega_r}{2Q \Delta\omega^2_1}.
\end{equation}
Substitute Eqs.~(\ref{eq:labe21}) and~(\ref{eq:labe22}) into Eq.~(\ref{eq:labe19}), the growth rate $\Omega_i$ of the D-mode instability can be obtained:
\begin{equation}\label{eq:labe23}
   \Omega_i=\frac{b-2\Omega_r/\tau_z}{2\Omega_r-k}.
\end{equation}
Please note that both $k$ and $b$ are positive. For the D-mode instability, we can find latter that $2\Omega_r<k$. Therefore, the D-mode instability is naturally damped by the PSHC fundamental impedance, while it is anti-damped by the radiation-damping effect. 

For the case of a relatively large detuning of PSHC and a realistic radiation damping time, it will be shown later that $2\Omega_r\ll k$ and $2\Omega_r/\tau_z \ll b$. In this case, we can get
\begin{equation}\label{eq:labe24}
   \Omega_i \approx -\frac{b}{k}=-\frac{\omega_r}{2Q}.
\end{equation}
It can be found that the D-mode damping rate is exactly equal to the voltage decaying rate of the PSHC (in other words, the cavity half bandwidth), which was not easily revealed by the search algorithm for the minimum proposed in Ref.~\cite{He06}.

\subsection{\label{sec:level3.3} Verification using the HALF parameters}
The Hefei Advanced Light Facility (HALF)~\cite{Bai13} is being constructed to be a 4th generation synchrotron light source, with a beam energy of $\SI{2.2}{GeV}$ and a nominal current of $\SI{350}{mA}$. To mitigate the Intrabeam scattering and Touschek scattering effects, a passive superconducting harmonic cavity will be adopted to stretch the bunch in the longitudinal direction. The HALF main parameters as well as the PSHC parameters, as listed in Table~\ref{tab:parameters}, will be used for verification of the aforementioned analytical formulas for the D-mode instability. 
\begin{table}[!hbt]
\setlength{\tabcolsep}{1.8mm}
   \centering
   \caption{Main parameters of the HALF storage ring used for the following calculations.}
   \begin{tabular}{lcc}
       \toprule
       \textbf{Parameter}&{Symbol}                  &{Value}               \\
       \hline
           Beam energy & $E$              & \SI{2.2}{GeV}          \\
           Ring circumference  & $C$           & \SI{479.86}{m}         \\
           Assumed beam current & $I_0$ &\SI{40}{mA}         \\
           Longitudinal damping time & $\tau_z$       & \SI{14}{ms}      \\
           Momentum compaction&
           $\alpha_c$
           & $9.4\times10^{-5}$           \\
           Harmonic number & $h$          & \SI{800}{}         \\
           Energy loss per turn &
           $U_0$
           & \SI{400}{keV}         \\
           Main cavity voltage&
           $V_{rf}$
           & \SI{1.2}{MV}        \\
           HHC harmonic order 
           & $n$             & \SI{3}{}        \\
           HHC normalized shunt impedance & $R/Q$              & \SI{39}{\Omega}        \\
           HHC quality factor &
           $Q$
           & $2\times10^8$          \\
           HHC near-optimum detuning & $\Delta f_r$              & \SI{6}{kHz}        \\
       \hline
   \end{tabular}
   \label{tab:parameters}
\end{table}

\begin{figure}[!htbp]  
  \centering
      \includegraphics[width=0.45\textwidth]{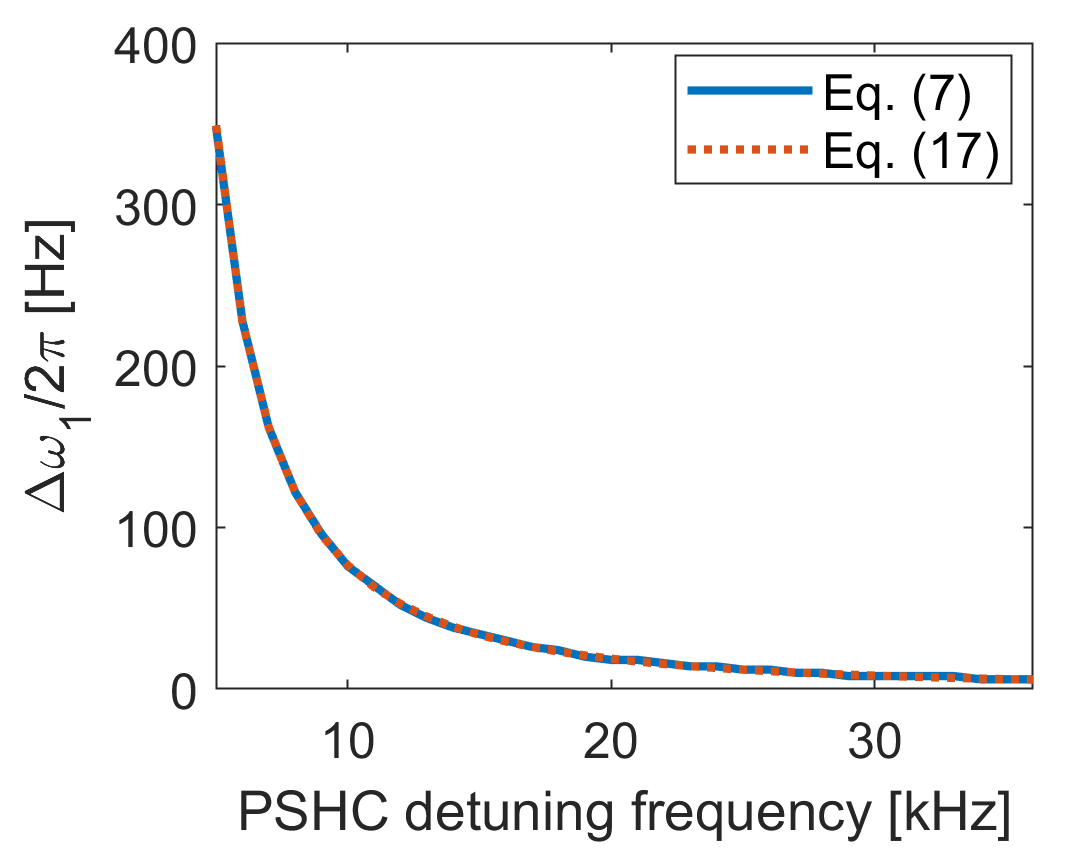}
  \caption{The frequency deviation $\Delta\omega_1/2\pi$ as functions of the PSHC detuning, obtained by solving directly Eq.~(\ref{eq:labe7}) and using Eq.~(\ref{eq:labe17}), respectively.}
  \label{fig1}
\end{figure}

\begin{figure}[!htbp]  
  \centering
      \includegraphics[width=0.45\textwidth]{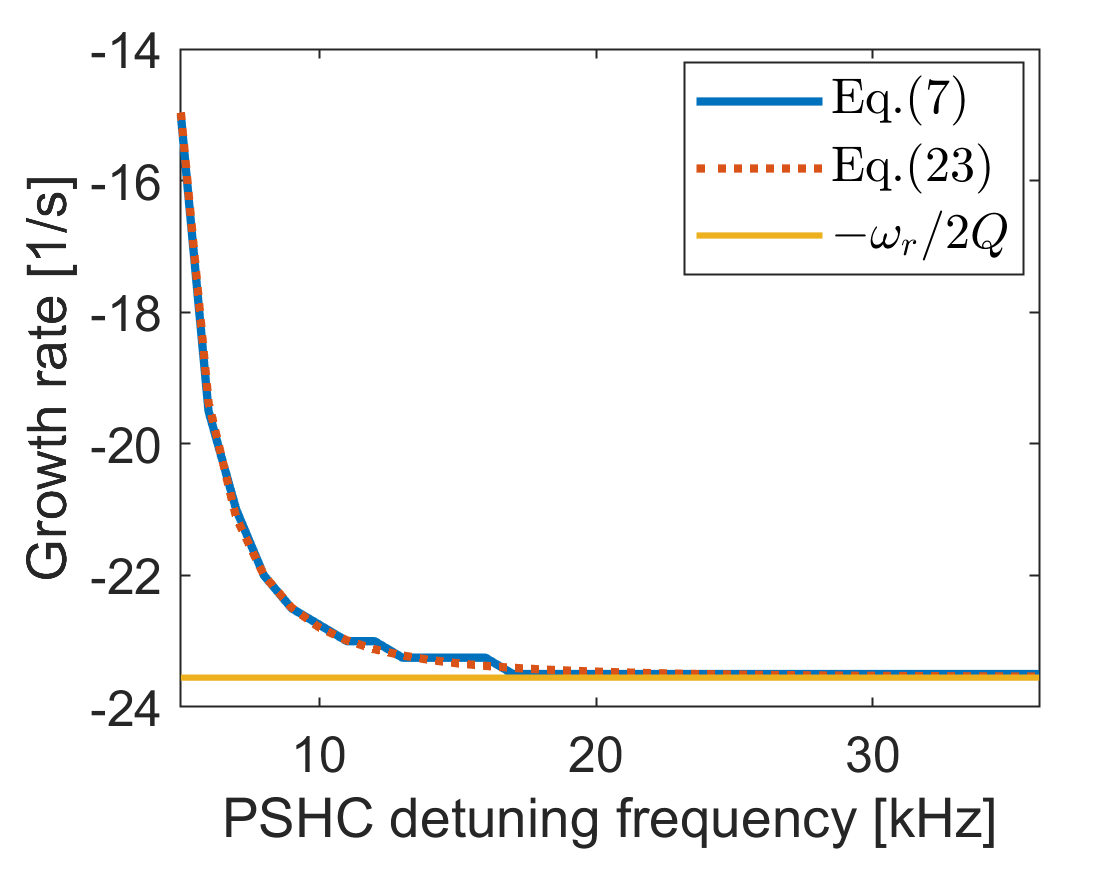}
  \caption{The D-mode growth rate as a function of the PSCH detuning, obtained by solving directly Eq.~(\ref{eq:labe7}) and using the analytical formula of Eq.~(\ref{eq:labe23}), respectively.}
  \label{fig2}
\end{figure}

Note that we chose a relatively low current of $\SI{40}{mA}$ rather than the nominal current as a study case. This case has already been studied in Ref.[6], mainly to reveal the characteristics of the D-mode instability. In this paper, we will prove that the approximate analytical formula is accurate and feasible for the D-mode instability by comparing it with the approach of solving directly the Robinson instability equation. Figure~\ref{fig1} shows the value of $\Delta\omega_1/2\pi$ (the frequency deviation of the PSHC detuning from the D-mode oscillation) as functions of the PSHC detuning. Figure~\ref{fig2} shows the corresponding growth rates. It can be seen that the approximate analytical formulas are in good agreement with the results obtained by directly solving the Robinson equation. Additionally, it can be seen from Fig.~\ref{fig2} that the D-mode damping rate approaches the cavity half bandwidth when increasing the PSHC detuning. It indicates that the formula of Eq.~(\ref{eq:labe24}) is correct. 

\begin{figure}[!htbp]  
  \centering
      \includegraphics[width=0.45\textwidth]{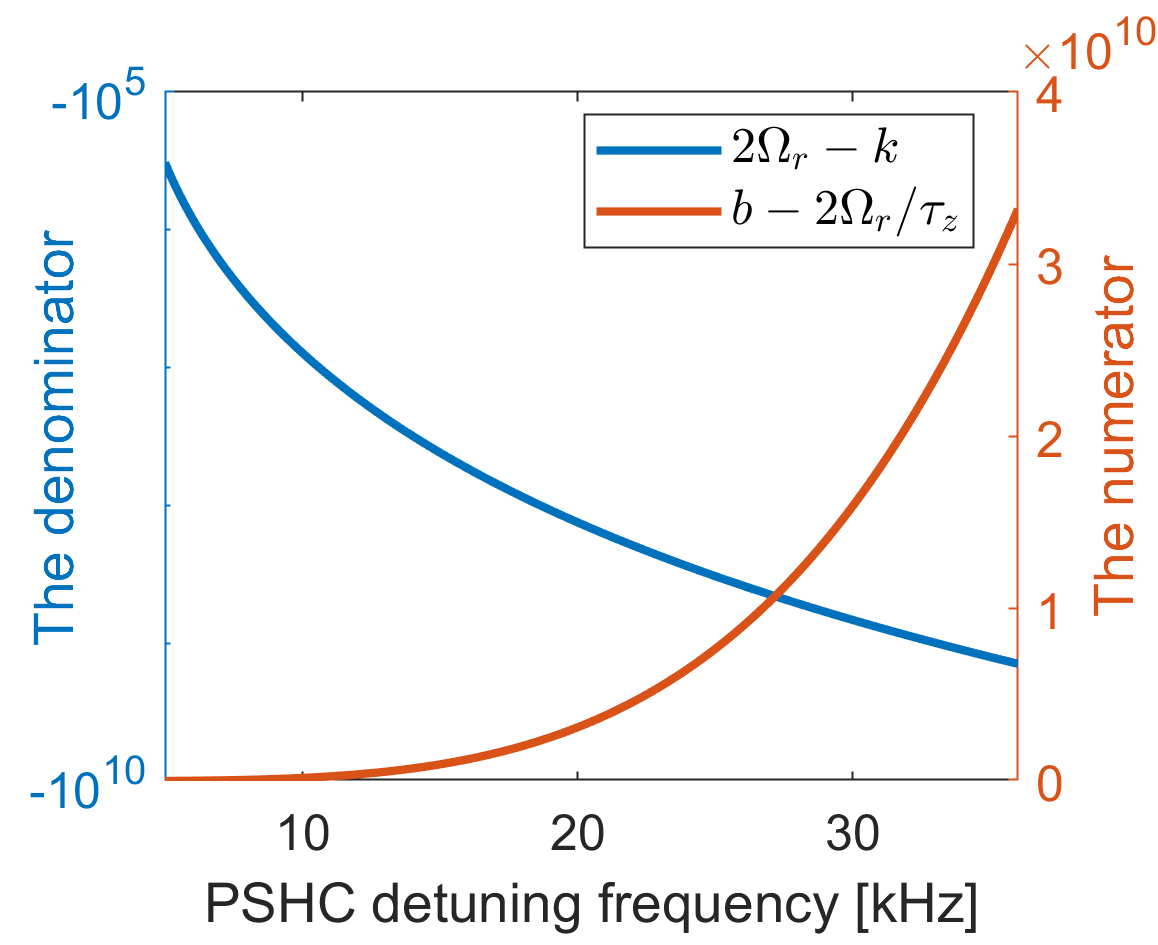}
  \caption{The denominator (in blue curve) and the numerator (in red curve) of Eq.~(\ref{eq:labe23}) as functions of the PSHC detuning frequency.}
  \label{fig3}
\end{figure}

The coefficients of $b$ and $k$ are inversely proportional to the square of $\Delta\omega_1$, as seen in Eqs.~(\ref{eq:labe21}) and~(\ref{eq:labe22}). Both of them affect the D-mode growth rate, as seen in Eq.~(\ref{eq:labe23}), where $b$ affects the numerator and $k$ affects the denominator. Figure~\ref{fig3} shows the numerator $b-2\Omega_r/\tau_z$  and denominator $k-2\Omega_r$ as functions of the PSHC detuning. As can be seen, the denominator is always less than zero, while the numerator approaches zero when reducing the PSHC detuning. Therefore, it inspires us that the growth or damping of the D-mode instability is entirely determined by the positive or negative value of the numerator, in other words, the D-mode stability relies on the competition of the PSHC fundamental mode damping against the radiation anti-damping. In the next section, we will show how to find the approximate formula of threshold detuning for the D-mode instability by solving the Equation $b-2\Omega_r/\tau_z =0$.

\section{\label{sec:level4} Threshold detuning for the D-mode instability}
\subsection{\label{sec:level4.1} Threshold detuning}
In Eq.~(\ref{eq:labe23}), as aforementioned, the denominator is generally less than zero for the D mode of interest. Therefore, the damp-grow of the D-mode oscillation is fully determined by the positive-negative of the numerator, and the threshold detuning can be obtained by zeroing the numerator $b-2\Omega_r/\tau_z =0$.  As expressed in Eq.~(\ref{eq:labe21}), $b$ is a function of $\Delta\omega_1$. Take $\Omega_r=\Delta\omega_r-\Delta\omega_1$, and Substitute it and Eq.~(\ref{eq:labe21}), we can get 
\begin{equation}\label{eq:labe25}
   2(\Delta\omega_r-\Delta\omega_1)=\frac{c R\tau_z\omega^2_r}{4Q^2\Delta\omega^2_1}.
\end{equation}
Considering $\Delta\omega_1 \ll \Delta\omega_r$, the second term on the left-hand side of Eq.~(\ref{eq:labe25}) can be safely neglected. Equation~(\ref{eq:labe25}) is thus simplified as
\begin{equation}\label{eq:labe26}
   \Delta\omega_r\Delta\omega^2_1=\frac{c R\tau_z\omega^2_r}{8Q^2}.
\end{equation}
The right-hand side represents a constant for a given beam current, which is independent of the PSHC detuning. The left-hand side is a function of the detuning. If given the detuning, its value can be easily calculated. Accurately solving Eq.~(\ref{eq:labe26}) is not a difficult task. The resulting detuning is the detuning threshold of the D mode for a given beam current.

For the convenience of solving Eq.~(\ref{eq:labe26}), the range of the solution can be predetermined, that is, the approximate solution can be obtained first, and the exact solution can thus be found around the approximate one. In Appendix B, it shows that $\Delta\omega_1\approx \frac{c\omega_rR}{2\Delta\omega^2_r Q}$. Substitute it into the left-hand side of Eq.(\ref{eq:labe26}), it finally gives
\begin{equation}\label{eq:labe27}
   \Delta\omega^{th}_r \approx \sqrt[3]{2I_0 \alpha_c \omega_r R/(T_0 \tau_z E/e)}=\eta_1 \cdot \sqrt[3]{I_0},
\end{equation}
where $\eta_1$ is defined as
\begin{equation}\label{eq:labe28}
   \eta_1=\sqrt[3]{2 \alpha_c \omega_r R/(T_0 \tau_z E/e)}.
\end{equation}
The D-mode threshold detuning is approximately proportional to the cubic root of the beam current. 

For PSHC, the actual detuning should be set to meet the specific requirements of cavity voltage for reaching the desired bunch lengthening. To facilitate the following discussion, we focus on the near-optimum bunch lengthening condition, with only setting the PSHC voltage amplitude at the optimum lengthening point (due to the non-optimum phase, it is generally called near-optimum). In that case, the PSHC voltage amplitude can be expressed as:
\begin{equation}\label{eq:labe29}
   V^{opt}_h=k_{opt}V_{rf}=\frac{F_{opt}I_0 \omega_r R}{\Delta\omega_r Q},
\end{equation}
where $k_{opt}$ is the corresponding ratio of the PSHC voltage and the main cavity voltage, and $F_{opt}$ is the corresponding bunch form factor amplitude. Generally, $k_{opt}$ can be pre-given according to the near-optimum lengthening condition, as well as $V^{opt}_h$ and $F_{opt}$. As a result, the PSHC detuning at the near-optimum lengthening point can be computed by:
\begin{equation}\label{eq:labe30}
   \Delta\omega^{opt}_r= \frac{F_{opt}I_0 \omega_r R}{V^{opt}_h Q}=\eta_2 \cdot I_0,
\end{equation}
where $\eta_2$ is given as:
\begin{equation}\label{eq:labe31}
   \eta_2= \frac{F_{opt} \omega_r R}{V^{opt}_h Q},
\end{equation}
Obviously, under the near-optimum PSHC voltage setting, the detuning is proportional to the beam current. 

\subsection{\label{sec:level4.2} Application to realistic storage rings}

\begin{figure*}[!tbh]
   \centering
   \includegraphics*[scale=0.64]{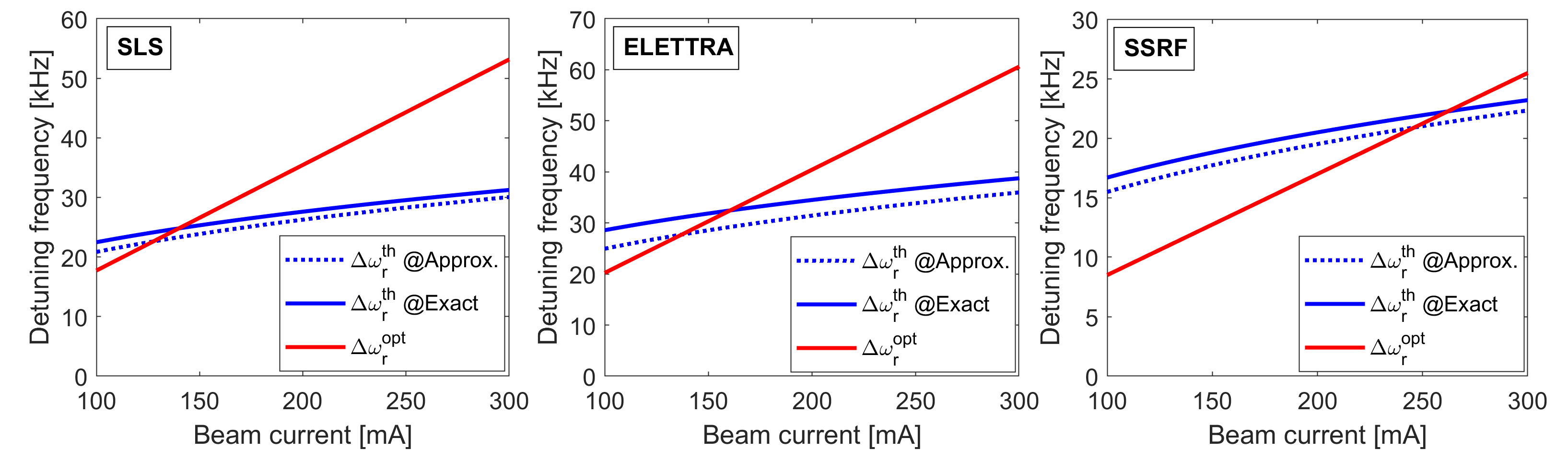}
   \caption{The D-mode threshold detuning (in blue curves) and the detuning required for near-optimum bunch lengthening (in red curves) as functions of the beam current. The solid blue curve is the exact solution and the dashed one is the approximate solution. The three plots on the left, middle, and right correspond to SLS, ELETRA, and SSRF, respectively. The D mode is stable when the red curve is above the blue. Otherwise, it is unstable.}
   \label{fig4} 
\end{figure*}

In this section, we will show how to use the above equations to evaluate the limitation of the D-mode instability on bunch lengthening for existing storage rings that employed PSHC, including SLS~\cite{SLS14}, ELETTRA~\cite{ELETTRA15, Penco16}, and SSRF~\cite{Wu17}. Table~\ref{tab:parameters2} summarizes their main parameters related to the threshold detuning and bunch lengthening calculations, as well as the corresponding coefficients of $\eta_1$ and $\eta_2$. Using the value of $\eta_1$ and Eq.~(\ref{eq:labe27}), we can easily calculate the approximate threshold detuning for a given beam current. Then we solve Eq.~(\ref{eq:labe26}) to obtain the exact threshold detuning around this approximate one. Figure~\ref{fig4} shows the resulting threshold detuning and the desired near-optimum detuning as a function of the average beam current. It can be seen that the approximate solution is generally lower than the exact one, and the deviation between them is relatively larger for ELETTRA, reaching a maximum of $\SI{3.6}{kHz}$ at the beam current $\SI{100}{mA}$. More importantly, it can be seen that there is an intersection between the near-optimum detuning curve and the threshold detuning curve. The current at this intersection is the D-mode threshold current under the near-optimum lengthening condition, which is denoted as $I_D$. The D-mode instability will occur when below that current.

\begin{table*}[!hbt]
\setlength{\tabcolsep}{3.9mm}
   \centering
   \caption{Summary of main parameters of SLS, ELETTRA, and SSRF storage rings.}
   \begin{tabular}{lccccc}
       \toprule
       \textbf{Parameters} &{Unit} &{SLS}  &{ELETTRA}  &{SSRF}\\
       \hline
           Beam energy & GeV & 2.4 & 2.0 & 3.5\\
           Circumference & m & 288 & 259.2 & 432\\
           Harmonic number & - & 480 & 432 & 720\\
           Longitudinal damping time & ms & 4.5 & 8 & 3\\
           Momentum compaction & - & $7\times10^{-4}$ & $16\times10^{-4}$ & $4.2\times10^{-4}$\\
           RMS bunch energy spread & - & $9.0\times10^{-4}$ & $8.0\times10^{-4}$ & $11.1\times10^{-4}$\\
           Energy loss per turn & keV & 600 & 256 & 1700\\
           Main cavity voltage & MV & 2.08 & 1.7 & 4.5\\
           PSHC resonant frequency & MHz & $\sim$1500 & $\sim$1500 & $\sim$1500\\
           PSHC $R/Q$ & $\Omega$ & 88.4 & 88.4 & 88\\
           PSHC $Q$ & - & $2\times10^8$ & $2\times10^8$ & $2\times10^8$\\
           PSHC $V^{opt}_h$ & kV & 660 & 559 & 1374\\
           Bunch form factor $F_{opt}$ & - & 0.883 & 0.853 & 0.885\\
           $\eta_1$ in Eq.~(\ref{eq:labe28}) & - & $2.82\times10^5$ & $3.72\times10^5$ & $2.10\times10^5$\\
           $\eta_1$ in Eq.~(\ref{eq:labe31})  & - & $1.11\times10^6$ & $1.27\times10^6$ & $5.34\times10^5$\\
           $\eta_1/\eta_2$ & - & 0.254 & 0.293 & 0.393\\
           Approximate $I_D=\sqrt[1.5]{\eta_1/\eta_2}$ & A & 0.128 & 0.137 & 0.246\\
           Exact $I_D$ & A & 0.140 & 0.160 & 0.262\\
           
       \hline
   \end{tabular}
   \label{tab:parameters2}
\end{table*}

For SLS, ELETTRA, and SSRF, their $I_D$ obtained for both the approximate and exact cases, are also listed in Table~\ref{tab:parameters2}. The approximate one is less than the exact one by 12, 23, and 16 mA for SLS, ELETTRA, and SSRF, respectively. The exact threshold current under the near-optimum lengthening condition is $\SI{160}{mA}$ for ELETTRA.  When the beam current gets lower than that, the detuning for the desired bunch lengthening becomes smaller than the D-mode threshold detuning. It indicates that the D-mode instability would occur before reaching the detuning for desired bunch lengthening. As a result, the bunch lengthening would be limited by the D-mode instability. We note that compared to SLS and ELETTRA, SSRF has a significantly larger threshold current. From numerical analysis, it is mainly due to its highest required cavity voltage. 

Compared with the 3rd generation synchrotron light sources including the aforementioned three ones, the 4th generation ones have a smaller momentum compaction factor, resulting in a significant reduction in the required main cavity voltage, as well as the harmonic cavity voltage for bunch lengthening. In addition, the radiation damping time is considerably larger. Both will alleviate the limitations of D mode on bunch lengthening. Taking HALF as an example, its $\eta_1/\eta_2 =0.0707$, corresponding to $I_D=18.8$ mA, indicating that HALF can avoid the D-mode instability for the near-optimum bunch lengthening at relatively very low currents. While, at 20mA, solving directly the Robinson equation shows that the HALF bunch lengthening will be limited by the S-D mode coupling. The S-D coupling case has been discussed in other papers~\cite{He06, Gamelin18}, which is beyond the scope of this paper. Nevertheless, the 4th generation rings generally have a relatively significantly lower D-mode threshold current compared to the 3rd generation ones.

\section{\label{sec:level5} Conclusion and discussion}
As a follow-up to our previous research [6], we have made further efforts in this paper to derive simpler analytical formulas for analyzing the D-mode instability. The contributions of this paper can be summarized as follows: (i) Based on the characteristic of D mode that it has a small frequency deviation from the PSHC detuning, appropriate approximations are taken to derive analytical formulas for calculating the D-mode frequency and growth rate, which makes its dependency on relevant parameters more intuitive. (ii) For the case of a relatively large detuning, the D-mode has a negative growth rate (is damped) whose absolute value approaches the PSHC half bandwidth. This has not been discovered in our previous study. (iii) We derived an approximate formula for calculating the D-mode threshold detuning, which indicates that the threshold detuning is approximately proportional to the cubic root of the beam current. Based on the approximate solution, it is easy to obtain the exact threshold detuning by solving Eq.~(\ref{eq:labe26}), which can be used to evaluate the limitation of the D-mode instability on bunch lengthening. (iv) It can be seen from the derivation process for the D-mode analytical formulas of Eqs.~(\ref{eq:labe18}) and~(\ref{eq:labe23}), the real part of the PSHC fundamental mode impedance mainly affects the value of $b$, while the imaginary part mainly affects $k$ and the D-mode frequency. Therefore, the imaginary part plays a more crucial role in the D-mode motion. 

It should be pointed out that the derivation method demonstrated in this paper is not limited to analyzing the D-mode instability but can also be extended to mode 1 instability~\cite{Marco19,He20}. However, this part is beyond the scope of this paper and will be presented in another paper.

Finally, it should be emphasized that the D-mode instability needs to be considered only when the PSHC is operated in bunch lengthening mode. For the case of bunch shortening, we can directly solve the mode-zero Robinson instability equation, and of course, we can also derive analytical formulas for analyzing the D-mode instability. In that case, it can be found that the D mode is always stable. The relevant derivations are shown in Appendix B for the reader's reference.

\begin{acknowledgments}
This research was supported by the National Natural Science Foundation of China (No. 12105284 and 12375324).
\end{acknowledgments}

\appendix
\renewcommand{\appendixname}{APPENDIX~}
\section{\label{sec:AppdenxA} Approximation of $\Delta\omega_1$}
\begin{figure}[!tbh]
   \centering
   \includegraphics*[scale=0.8]{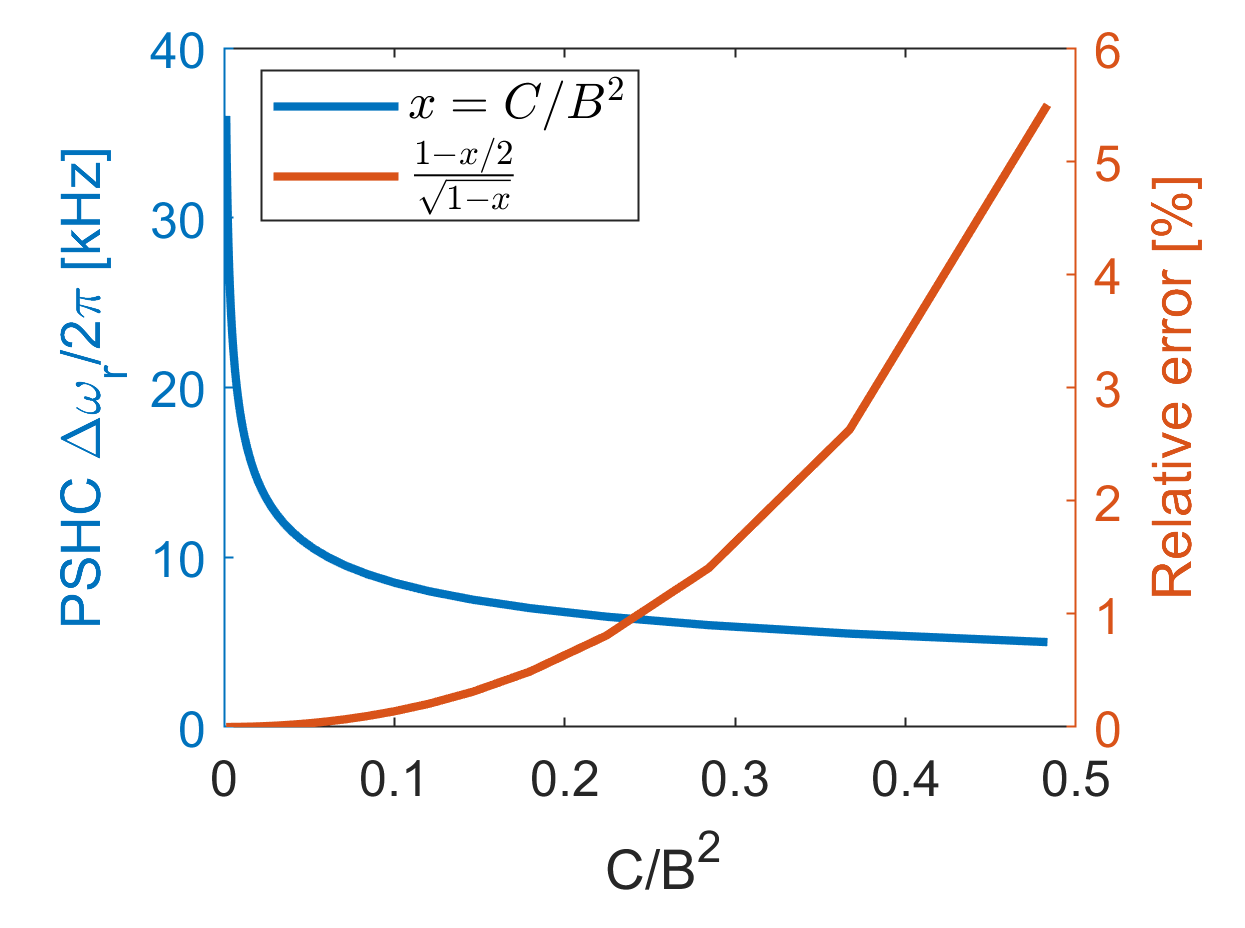}
   \caption{The support for the approximation from Eq.~(\ref{eq:labeA1}) to Eq.~(\ref{eq:labeA2}).}
   \label{fig5}
\end{figure}

\begin{figure}[!tbh]
   \centering
   \includegraphics*[scale=0.85]{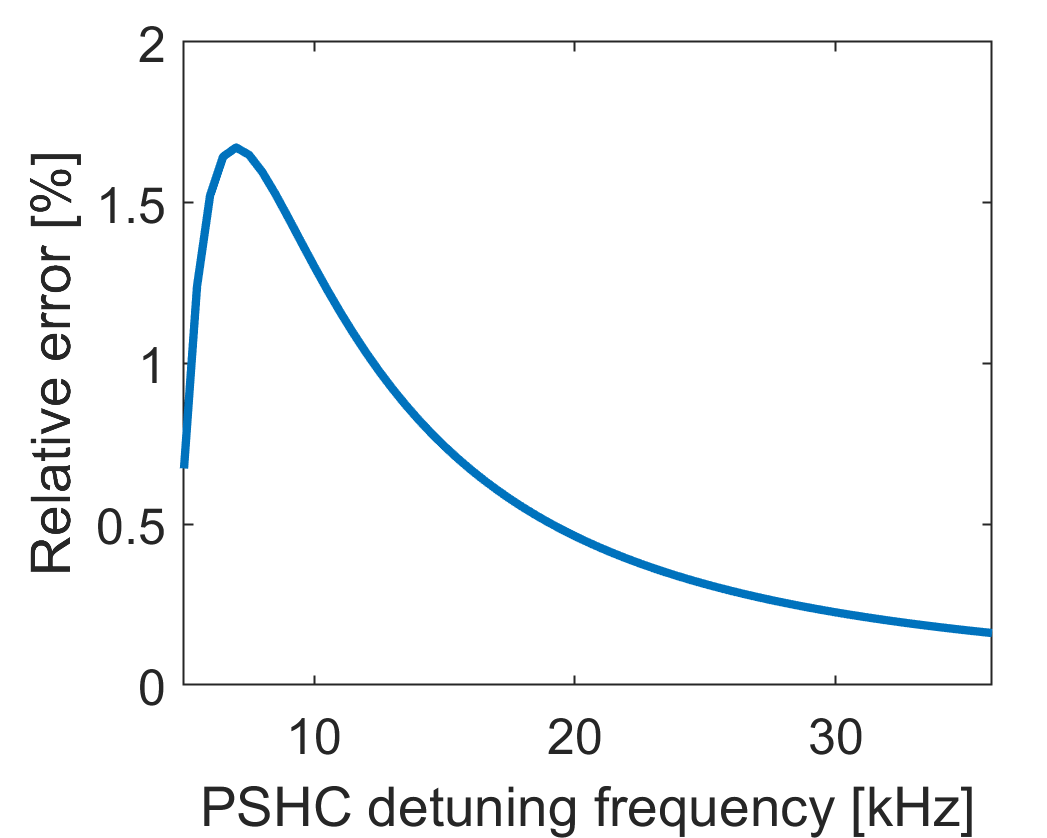}
   \caption{The relative error between the approximate and exact values of $B$.}
   \label{fig6}
\end{figure}

\begin{figure}[!tbh]
   \centering
   \includegraphics*[scale=0.85]{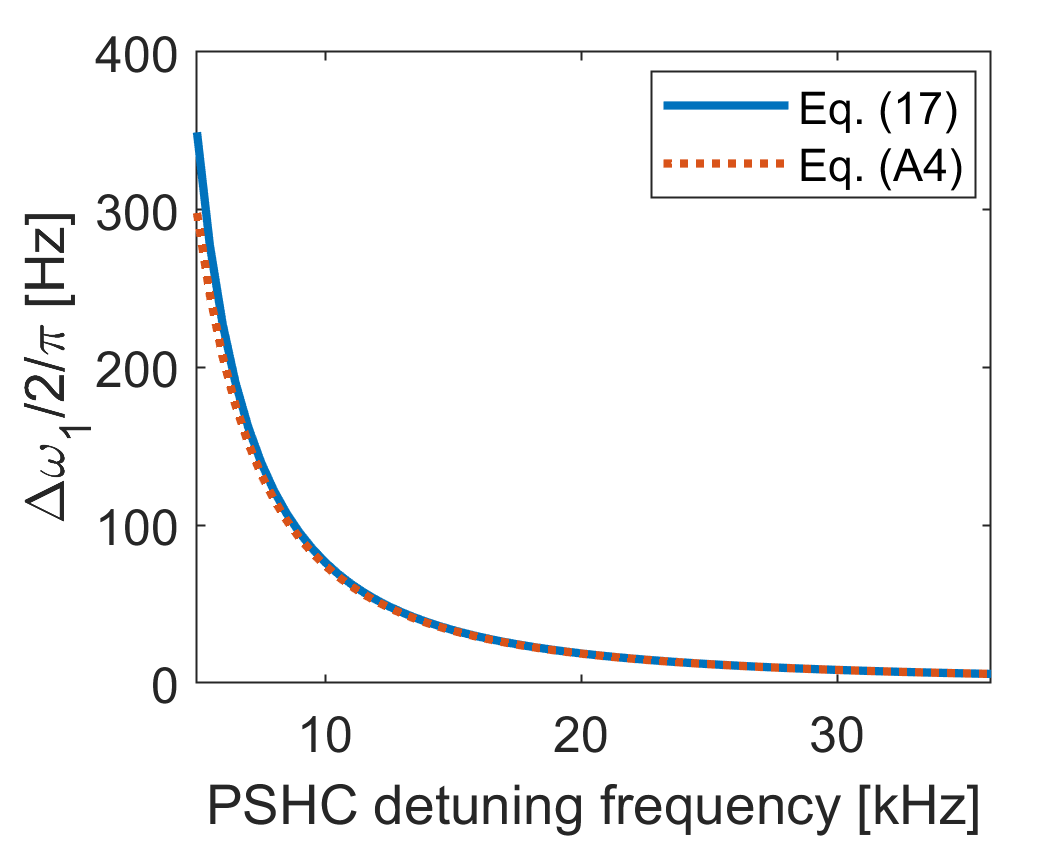}
   \caption{The comparison of the exact and the approximate values of $\Delta\omega_1$, computed by Eq.~(\ref{eq:labe17}) and Eq.~(\ref{eq:labeA4}), respectively}
   \label{fig7}
\end{figure}

Here we rewrite the formula of $\Delta\omega_1$ as follows
\begin{equation}\label{eq:labeA1}
   \Delta\omega_1=B-B\sqrt{1-C/B^2},
\end{equation}
where $B$ and $C$ are determined by Eqs.~(\ref{eq:labe15}) and~(\ref{eq:labe16}), respectively. Still using the HALF parameters, the resulting $C/B^2$ as a function of the PSHC detuning is shown in Fig.~\ref{fig5}. It can be seen that the maximum of $C/B^2$ is lower than 0.5 at the lowest detuning of $\SI{5}{kHz}$. Figure~\ref{fig5} also shows that the relative error between $1-C/2B^2$ and $\sqrt{1-C/B^2}$ is less than 6$\%$. It indicates that the approximation of $\sqrt{1-C/B^2}\approx 1-C/2B^2$ is reasonable because $C/B^2 <0.5$. Then Eq.~(\ref{eq:labeA1}) can be simplified as:
\begin{equation}\label{eq:labeA2}
   \Delta\omega_1 \approx \frac{C}{2B},
\end{equation}
The expression of $B$ can also be simplified as:
\begin{equation}\label{eq:labeA3}
   B=\frac{\Delta\omega_r}{4}-\frac{\omega^2_s}{4\Delta\omega_r}-\frac{cR\omega_r}{16Q\Delta\omega^2_r}\approx \frac{\Delta\omega_r}{4}.
\end{equation}
where the relative error between the approximate and the exact values of $B$ is shown in Fig.~\ref{fig6}. The maximum relative error is less than 2$\%$. Therefore, the above approximation used in Eq.~(\ref{eq:labeA3}) is also reasonable.
Substitute Eq.~(\ref{eq:labeA3}) and Eq.~(\ref{eq:labe16}) into Eq.~(\ref{eq:labeA2}), we finally get:
\begin{equation}\label{eq:labeA4}
   \Delta\omega_1 \approx \frac{c\omega_r R}{2\Delta\omega^2_r Q},
\end{equation}
Figure~\ref{fig7} shows the exact and the approximate values of $\Delta\omega_1$, obtained using Eq.~(\ref{eq:labe17}) and Eq.~(\ref{eq:labeA4}), respectively. It can be seen that Eq.~(\ref{eq:labeA4}) is in good agreement with Eq.~(\ref{eq:labe17}), especially at a relatively large detuning of PSHC.

\section{\label{sec:AppdenxB} Analytical formulas for the D mode in the case of bunch shortening driven by PSHC}
\begin{figure}[!tbh]
   \centering
   \includegraphics*[scale=0.35]{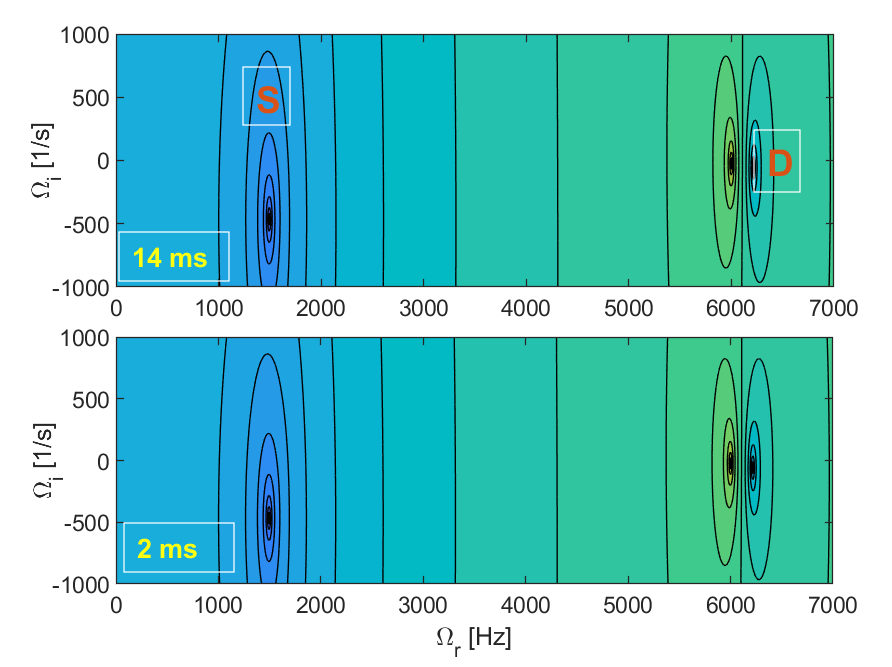}
   \caption{The contour map of log$(|g(\Omega)|)$ for the case of detuning of $-6$ kHz. The top and the bottom correspond to the cases of damping times of 14 ms and 2 ms, respectively. The left local minimum corresponds to the S mode, while the right one to the D mode.}
   \label{fig8}
\end{figure}

For the case of PSHC operating in bunch shortening mode, the search algorithm for minimum can be also used directly to solve the mode-zero Robinson instability equation. We still take the HALF storage ring as an example. Using the parameters listed in Table~\ref{tab:parameters} and assuming the detuning frequency of $-6$ kHz, the resulting 2D contour maps of log$(|g(\Omega)|)$ are shown in Fig.~\ref{fig8}. It can be seen that for both cases of $\tau_z$=14 ms and $\tau_z$=2 ms, the D-mode frequency is slightly larger than the detuning of PSHC, which is opposite to the case of bunch lengthening. This feature can be utilized to derive the oscillation frequency and growth rate formulas for the D mode.

Take the real part of Eq.~(\ref{eq:labe8}), and ignore unimportant terms, it also gives
\begin{equation}\label{eq:labeB1}
      \Omega^2_r-\omega^2_s+c \cdot \mathrm {Im}\{Z(\omega^{+}_p)-Z(\omega^{-}_p)\}= 0.
\end{equation}
Here, we set $\Delta\omega_r=nh\omega_0-\omega_r>0$,  and it is already known that $\Omega_r$ is slightly higher than $\Delta\omega_r$. If introduce $\Delta\omega_1=\Omega_r+\Delta\omega_r$ and $\Delta\omega_2=\Omega_r-\Delta\omega_r$, the third term of Eq.~(\ref{eq:labeB1}) can be transformed as follows:
\begin{equation}\label{eq:labeB2}
   \begin{aligned}
   c \cdot \mathrm {Im}\{Z(\omega^{+}_p)&-Z(\omega^{-}_p)\}=c \cdot \mathrm {Im}\{Z(\omega^{+}_p)+Z(-\omega^{-}_p)\}\\
   &=c \cdot \mathrm {Im}\{\frac{R}{1-i\frac{2Q\Delta\omega_1}{\omega_r}}+\frac{R}{1+i\frac{2Q\Delta\omega_2}{\omega_r}}\}\\
   &\approx \frac{cR\omega_r}{4Q\Delta\omega_r}-\frac{cR\omega_r}{2Q\Delta\omega_2}.
   \end{aligned}
\end{equation}
Substitute Eq.~(\ref{eq:labeB2}) into Eq.~(\ref{eq:labeB1}), and take $\Omega_r=\Delta\omega_r+\Delta\omega_2$, it gives:
\begin{equation}\label{eq:labeB3}
   \Delta\omega^3_2+2\Delta\omega_r \Delta\omega^2_2+(\Delta\omega^2_r-\omega^2_s+\frac{cR\omega_r}{4Q\Delta\omega_r})\Delta\omega_2-\frac{c\omega_r R}{2Q}=0.
\end{equation}
Equation~(\ref{eq:labeB3}) is a quadratic equation, which can be solved directly to obtain $\Delta\omega_2$.

Actually, due to $\Delta\omega_2\ll \Delta\omega_r$, the first terms of Eq.~(\ref{eq:labeB3}) can be omitted. Equation~(\ref{eq:labeB3}) can be further simplified as:
\begin{equation}\label{eq:labeB4}
   \Delta\omega^2_2+(\frac{\Delta\omega_r}{2}-\frac{\omega^2_s}{2\Delta\omega_r}+\frac{cR\omega_r}{8Q\Delta\omega^2_r})\Delta\omega_2-\frac{c\omega_r R}{4\Delta\omega_rQ}=0.
\end{equation}
If we introduce
\begin{equation}\label{eq:labeB5}
   B=\frac{\Delta\omega_r}{4}-\frac{\omega^2_s}{4\Delta\omega_r}+\frac{cR\omega_r}{16Q\Delta\omega^2_r}.
\end{equation}
and
\begin{equation}\label{eq:labeB6}
   C= \frac{c\omega_r R}{4\Delta\omega_r Q},
\end{equation}
it is easy to know Eq.~(\ref{eq:labeB4}) has two solutions of $-B\pm\sqrt{B^2+C}$. While only one closet to zero is required. Therefore, we get
\begin{equation}\label{eq:labeB7}
   \Delta\omega_2=-B+\sqrt{B^2+C},
\end{equation}
The D-mode frequency can be written as:
\begin{equation}\label{eq:labeB8}
   \Omega_r=\Delta\omega_r+\Delta\omega_2,
\end{equation}
Take the imaginary part of Eq.~(\ref{eq:labe8}), it gives:
\begin{equation}\label{eq:labeB9}
   2\Omega_r\Omega_i+\frac{2\Omega_r}{\tau_z}-c \cdot \mathrm {Re}\{Z(\omega^{+}_p)-Z(\omega^{-}_p)\}=0.
\end{equation}
Using $\mathrm {Re}\{Z(\omega^{-}_p)\}=\mathrm {Re}\{Z(-\omega^{-}_p)\}$, the third terms of Eq.~(\ref{eq:labeB9}) can be transformed as follows:
\begin{equation}\label{eq:labeB10}
   \begin{aligned}
   c \cdot \mathrm {Re}\{Z(\omega^{+}_p)&-Z(\omega^{-}_p)\} =c \cdot \mathrm {Re}\{Z(\omega^{+}_p)-Z(-\omega^{-}_p)\}\\
   &=c \cdot \mathrm {Re}\{\frac{R}{1+i\frac{2Q\Delta\Omega_1}{\omega_r}}-\frac{R}{1+i\frac{2Q\Delta\Omega_2}{\omega_r}}\}\\
   &\approx k\Omega_i+b.
   \end{aligned}
\end{equation}
where $\Delta\Omega_1=-\Delta\omega_r-\Omega_r-i\Omega_i=-\Delta\omega_1-i\Omega_i$, and $\Delta\Omega_2=-\Delta\omega_r+\Omega_r+i\Omega_i=\Delta\omega_2+i\Omega_i$.

To obtain $b$, let $\Omega_i=0$, and ignore the term $\mathrm {Re}\{Z(\omega^{+}_p)\}$ due to $\Delta\omega_2 \ll \Delta\omega_1$, then it gives
\begin{equation}\label{eq:labeB11}
   b|_{\Omega_i=0}\approx c \cdot \mathrm {Re} \{-\frac{R}{1+i\frac{2Q\Delta\omega_2}{\omega_r}}\} \approx -\frac{cR\omega^2_r}{4Q^2\Delta\omega^2_1}.
\end{equation}

To obtain $k$, we first get the derivative of $\mathrm {Re}\{Z(\omega^{+}_p)-Z(\omega^{-}_p)\}$ about $\Omega_i$, and let $\Omega_i=0$. The term $\mathrm {Re}\{Z(\omega^{+}_p)\}$ can also be omitted due to $\Delta\omega_2 \ll \Delta\omega_r$, then we get
\begin{equation}\label{eq:labeB12}
   k|_{\Omega_i=0}\approx \frac{2cQ}{\omega_r R} \mathrm {Re} \{\frac{R^2}{(1+i\frac{2Q\Delta\omega_2}{\omega_r})^2}\} \approx -\frac{cR\omega_r}{2Q \Delta\omega^2_2}.
\end{equation}
Substitute $b$ and $k$ into Eq.~(\ref{eq:labeB9}), we finally get the D-mode growth rate:
\begin{equation}\label{eq:labeB13}
   \Omega_i=\frac{b-2\Omega_r/\tau_z}{2\Omega_r-k}.
\end{equation}
It should be noted that both $k$ and $b$ are negative always. It indicates that $\Omega_i<0$, in other words, the D-mode oscillation is naturally damped for the case of bunch shortening. Both PSHC fundamental impedance and radiation damping contribute to the D-mode damping. 

When the detuning is relatively large, it has $-b\gg 2\Omega_r/\tau_z$  and $-k\gg 2\Omega_r$. The D-mode growth rate can be given as:
\begin{equation}\label{eq:labeB14}
   \Omega_i \approx -\frac{b}{k}=-\frac{\omega_r}{2Q},
\end{equation}
whose absolute value is exactly equal to the PSHC half bandwidth.

\nocite{*}

\end{document}